\advance\voffset by -20mm
\advance\hoffset by -15mm
\documentstyle[12pt]{article}
\begin{document}

\null \vspace{30mm}
\setlength{\baselineskip}{20pt}

\centerline{\LARGE {An alternative to relativistic transformation of }}
\centerline{\LARGE {special relativity based on the first principles}}

\vspace{1.5cm}

\centerline{Young-Sea Huang}

\centerline{Department of Physics, Soochow University, Shih-Lin,
            Taipei, Taiwan }  
\centerline{yshuang@mail.scu.edu.tw}

\vspace{1.5cm}

\begin{abstract}
A new relativistic transformation in the velocity space (here named the differential Lorentz transformation) is formulated solely from the principle of relativity and the invariance of the speed of light. The differential Lorentz  transformation is {\it via} transforming  physical quantities, instead of  space-time coordinates, to make laws of nature form-invariant.  The differential Lorentz transformation may provide a way to resolve the incompatibility of the theory of special relativity and the quantum theory. 

\end{abstract}

\vfill



\section{Introduction}

 The Lorentz transformation of space-time coordinates is the core of Einstein's special relativity. Many  derivations of the Lorentz transformation have been given.\cite{1}
 Most of those derivations are standard and more or less the same as the original method given by Einstein\cite{2}. A few of them are not so standard. For example, an alternative derivation has been presented using radar methods to determine the space and time coordinates of events -- though based on  Einstein's same two postulates, the principle of relativity and the invariance of the speed of light.\cite{3} More abstractly, there exist
derivations  based on hypotheses of the space-time structure and group properties.\cite{4}
 All the existing derivations start with an implicit assumption that the relativistic transformation is a global transformation of the space-time coordinates of events among inertial reference frames.

However, we think that relativistic transformations should be transformations of physical quantities (instead of  space-time coordinates) that make laws of nature form-invariant among inertial frames. Previously, with such a perspective on relativistic transformations,  a new relativistic transformation in the velocity space was derived based on  three assumptions: the principle of relativity, the invariance of the speed of light, and the transverse Doppler effect.\cite{5} The transverse Doppler effect is a consequence of the usual Lorentz transformation of space-time coordinates. In order to show that the new  relativistic transformation is independent of the usual Lorentz transformation, it is necessary to derive the  new relativistic transformation without assuming the transverse Doppler effect. 

It is interesting to note that the transverse Doppler effect is shown as a consequence of the new relativistic transformation (the differential Lorentz transformation), without involving the usual Lorentz transformation.\cite{6}
Furthermore, in studying the case that the medium moves at superluminal speeds  opposite to the propagation direction of plane waves, a remarkable finding is that plane waves propagate with negative frequencies, according to the usual Lorentz transformation  of space-time coordinates. However, physical waves can not propagate with negative frequencies. The negative frequency problem is resolved by the new relativistic transformation.\cite{7} Thus, this example explicitly indicates that the new relativistic transformation and the usual Lorentz transformation are not equivalent.
In the following, we will present the detailed derivation of the new relativistic transformation based on only the first-principles: the principle of relativity and the invariance of the speed of light.


\setcounter{page}{2}

\noindent

\section{Derivation of the new relativistic transformation based on the first-principles}

In order to ensure that transformations
of physical quantities among inertial reference frames are physically meaningful,
the coordinate systems of the reference frames
must have the same constructive stipulation of their space
and time units. All the coordinate systems of the inertial reference frames under
consideration are presumed
to have identical stipulations of their space and time measurements.\cite{8,9} 
Consider a particle moving in an inertial reference frame which has such a setup coordinate system. With respect to this frame, during an infinitesimal time interval $dt>0$,
the particle will move with a spatial displacement $d{\bf x}={\bf v}dt$,
where  ${\bf v}$ is the instantaneous velocity of the particle. Since the speed of light $c$ is an invariant constant, 
we define $dx^{0} \equiv c \,dt$. Then, we can use the four-vector of infinitesimal displacement
$d x^{\alpha}\, \equiv \, (dx^{0}, d{\bf x})$ to characterize
the state of motion of the particle. It should be noted that according to the uncertainty principle of quantum theory it is impossible to definitely know the position ${\bf x}$ of a particle when its velocity is exactly determined, that is, when its state of motion $d x^{\alpha}$ is exactly specified. The uncertainty principle does not stop one from merely assuming the particle is at some {\it unknown} position when its state of motion $d x^{\alpha}$ is exactly specified. However, the particle is at someplace that can not be determined without interfering its state of motion.  
It is not allowed to specify exact values for both the state of motion $d x^{\alpha}$ and the position ${\bf x}$ of a particle simultaneously. Here, to have the definite state of motion $d x^{\alpha}$, we need not (it is even impossible to) specify the definite values for $x^{\alpha}$.

    Suppose an inertial reference frame $\bar{S}$ moves with a constant velocity
${\bf V}$ with respect to another inertial reference frame $S$.  At an arbitrary instant of time, a particle has the state of motion $dx^{\alpha}$
with respect to the frame $S$ with the setup of Cartesian coordinate system $X$. The particle has
the corresponding state of motion $d\bar{x}^{\mu}$ with respect to the frame $\bar{S}$ with the setup of Cartesian coordinate system $\bar{X}$.
Assume that there exists a universal transformation on the evolving state of motion and that it is a linear transformation. 
That is,
\begin{equation}\label{eq1}
dx^{\alpha} = a^{\alpha}_{\mu}(X,\bar{X}) d\bar{x}^{\mu}\quad
(\alpha ,\mu =0,1,2,3).
 \end{equation}
The transformation is assumed to be linear, since particles in uniformly rectilinear motion with respect to an inertial frame must also be in uniformly rectilinear motion with respect to all other inertial frames.
The coefficients $a^{\alpha}_{\mu}(X,\bar{X})$  depend on the relationships between the coordinate system $X$ of the frame $S$ and the coordinate system $\bar{X}$ of
the frame $\bar{S}$, such as the orientation of coordinate systems and the relative
velocity ${\bf V}$ between the
reference frames, but not at all on the motion of the particle.

    In the following derivation, we simplify the mathematical problem by keeping the
corresponding coordinate axes of the frames $S$ and $\bar{S}$ parallel and by taking their 
relative velocity to be along the direction of a chosen axis. Suppose the frame $\bar{S}$ moves with the speed $V$ along the positive $X^{1}$-axis with respect to the frame $S$. The
coefficients $a^{\alpha}_{\mu}(X,\bar{X})$ is also denoted simply as $a^{\alpha}_{\mu}$.

\begin{enumerate}

\renewcommand{\labelenumi}{\Roman{enumi}. }

\item
 Suppose that a particle co-moves with the  frame $\bar{S}$. Then, we have $d\bar{x}^{1} = d\bar{x}^{2} = d\bar{x}^{3} = 0$ for this particle  with respect to the  frame $\bar{S}$. Also, we have $dx^{2}
= dx^{3} = 0$ and $dx^{1}/dx^{0} = V/c \equiv \beta$ for this particle with respect to the frame $S$. Therefore, from the
equation of linear transformation Eq.~(\ref{eq1}), we have $a^{2}_{0} = a^{3}_{0} =0$ and
$a^{1}_{0}/a^{0}_{0} = dx^{1}/dx^{0} = \beta$. We consider elapses of time to be  always positive, that is, $dx^{0}>0$ and $d\bar{x}^{0}>0$. Thus, we have $a^{0}_{0} >0$. 

\item
Suppose that a particle co-moves with the frame $S$. Then, we have $dx^{1} = dx^{2} = dx^{3} = 0$ for this particle  with respect to the  frame $S$. Given a velocity ${\bf V}$ of the  frame
$\bar{S}$ relative to the  frame $S$, it is assumed that measured relative to the
 frame $\bar{S}$, the  frame $S$ has velocity $-{\bf V}$. Thus, we have $d\bar{x}^{2} =
d\bar{x}^{3} = 0$ and $d\bar{x}^{1}/d\bar{x}^{0} = -V/c = -\beta$ for this particle with respect to the frame $\bar{S}$.
Then, from Eq.~(\ref{eq1}) we have $a^{2}_{1} = a^{3}_{1} =0$ and
$-a^{1}_{0}/a^{1}_{1} = d\bar{x}^{1}/d\bar{x}^{0} =  -\beta$. 
With the results above, we obtain $a^{1}_{1} = a^{0}_{0}$.

\item
If a particle moves in the $X^{1}-X^{2}$ plane, then
$dx^{3} = 0 $ and $d\bar{x}^{3} =0 $. Therefore,
we have $a^{3}_{2} = 0$.

\item
If a particle moves in the $X^{1}-X^{3}$ plane, then
$dx^{2} = 0 $ and $d\bar{x}^{2} =0$.
Therefore, we have $a^{2}_{3} = 0$.

\item
If a particle moves in the $X^{2}-X^{3}$ plane, then
$dx^{1} = 0 $, but  $dx^{2}/dx^{0}$ and $dx^{3}/dx^{0}$  are
arbitrary. Hence, with respect to the frame $\bar{S}$, the particle moves with a velocity component $d\bar{x}^{1}/d\bar{x}^{0} = -\beta$. The velocity components $d\bar{x}^{2}/d\bar{x}^{0}$ and
$d\bar{x}^{3}/d\bar{x}^{0}$ are arbitrary.
Therefore, we have $a^{1}_{2}(d\bar{x}^{2}/d\bar{x}^{0}) +
a^{1}_{3}(d\bar{x}^{3}/d\bar{x}^{0}) =0$.
Because $d\bar{x}^{2}/d\bar{x}^{0}$ and
$d\bar{x}^{3}/d\bar{x}^{0}$ are
arbitrary, $a^{1}_{2} = a^{1}_{3} = 0$.

\item
If a particle moves in the $\bar{X}^{2}-\bar{X}^{3}$ plane,
then $d\bar{x}^{1} = 0 $, but $d\bar{x}^{2}/d\bar{x}^{0}$ and
$d\bar{x}^{3}/d\bar{x}^{0}$ are
arbitrary. Hence, the particle moves with a velocity such that $dx^{1}/dx^{0} = \beta$, but  $dx^{2}/dx^{0}$ and $dx^{3}/dx^{0}$ are arbitrary, with respect to the frame $S$.
Therefore, we have $dx^{1}/dx^{0}
= a^{1}_{0}/(a^{0}_{0} + a^{0}_{2}(d\bar{x}^{2}/d\bar{x}^{0})
+ a^{0}_{3}(d\bar{x}^{3}/d\bar{x}^{0})) = \beta$.  Since $d\bar{x}^{2}/d\bar{x}^{0}$ and
$d\bar{x}^{3}/d\bar{x}^{0}$ are
arbitrary, we have $a^{0}_{2} = a^{0}_{3} = 0$. Thus, the
universal transformation reduces to
\begin{equation}\label{eq2}
\left\{ \begin{array}{rcl}
          dx^{0}\!\!\! & = & \, a^{0}_{0}d\bar{x}^{0}+a^{0}_{1}d\bar{x}^{1} \\
          dx^{1}\!\!\! & = & \, \beta \, a^{0}_{0}d\bar{x}^{0}+a^{0}_{0}d\bar{x}^{1} \\
          dx^{2}\!\!\! & = & \, a^{2}_{2}d\bar{x}^{2} \\
          dx^{3}\!\!\! & = & \, a^{3}_{3}d\bar{x}^{3}\,\,\, .
         \end{array} \right.
\end{equation}

\item
 Consider a light pulse traveling parallel to the positive  $\bar{X}^{2}$-axis with respect to the frame $\bar{S}$. Then, $d\bar{x}^{1} = 0 $, $d\bar{x}^{3} = 0 $, and $d\bar{x}^{2} > 0 $.
From Eq.~(\ref{eq2}), $dx^{0}= a^{0}_{0}d\bar{x}^{0}$, $dx^{1}= \beta \, a^{0}_{0}d\bar{x}^{0}$, $dx^{2}= a^{2}_{2}d\bar{x}^{2}$, and $dx^{3}= 0$. Since the light pulse also travels with the constant speed $c$ with respect to the frame $S$ (the second postulate), $(dx^{1})^{2}+ (dx^{2})^{2}+ (dx^{3})^{2}= (dx^{0})^{2}$. Consequently, we obtain $(a^{2}_{2})^{2} = (1-\beta^{2}) (a^{0}_{0})^{2}$. This equation implies that $1-\beta^{2}>0$, and thus $V<c$. That means the relative velocity between frames must be less than the speed of light. Also, $a^{2}_{2}>0$, because $d\bar{x}^{2} > 0 $ and $dx^{2} > 0 $.
Similarly, consider a light pulse traveling parallel to the positive  $\bar{X}^{3}$-axis with respect to the frame $\bar{S}$. We have $(a^{3}_{3})^{2} = (1-\beta^{2}) (a^{0}_{0})^{2}$, and $a^{3}_{3}>0$.
Thus, $a^{2}_{2} = a^{3}_{3}$.

\item
 Consider a light pulse traveling parallel to the positive  $\bar{X}^{1}$-axis with respect to the frame $\bar{S}$. Then, $d\bar{x}^{1} > 0 $, $d\bar{x}^{2} = 0 $, and $d\bar{x}^{3} = 0 $.
From Eq.~(\ref{eq2}), $dx^{0}= a^{0}_{0} d\bar{x}^{0} + a^{0}_{1} d\bar{x}^{1}$, $dx^{1}= \beta \, a^{0}_{0} d\bar{x}^{0} 
+ a^{0}_{0} d\bar{x}^{1}$, $dx^{2}= 0$,  and $dx^{3}= 0$. Since the light pulse also travels with the constant speed $c$ with respect to the frame $S$, $(dx^{1})^{2}+ (dx^{2})^{2}+ (dx^{3})^{2}= (dx^{0})^{2}$. Consequently, we obtain $(1+ {a^{0}_{1}/ a^{0}_{0}})^{2} = (1+\beta)^{2}$. Therefore, we have either $a^{0}_{1}/ a^{0}_{0}=\beta$, or $a^{0}_{1}/a^{0}_{0} =-2-\beta$. By substituting $a^{0}_{1}/ a^{0}_{0}=-2-\beta$ into $dx^{0}= a^{0}_{0}d\bar{x}^{0}+a^{0}_{1}d\bar{x}^{1}$, we obtain $dx^{0}<0$.  However, since the elapse of time must be positive, $a^{0}_{1}/ a^{0}_{0}=-2-\beta$ is unacceptable. Hence, we have only $a^{0}_{1}= \beta\, a^{0}_{0}$. Then, the
universal transformation reduces further to
\begin{equation}\label{eq3}
\left\{ \begin{array}{rcl}
          dx^{0}\!\!\! & = & \,k(V)\,\gamma (d\bar{x}^{0} + \beta \, d\bar{x}^{1}) \\
           dx^{1}\!\!\! & = & \,k(V)\,\gamma (\beta \, d\bar{x}^{0} + d\bar{x}^{1}) \\
          dx^{2}\!\!\! & = & \,k(V)\, d\bar{x}^{2}   \\
          dx^{3}\!\!\! & = &\,k(V)\, d\bar{x}^{3}\,\,\, ,
         \end{array} \right.
\end{equation}
 where $\gamma \equiv (1-\beta ^{2})^{-1/2}$, and $k(V)=a^{2}_{2} = a^{3}_{3}>0$.

\item
We have left only to determine the value of $k(V)$. From the point of view of the frame $\bar{S}$, the $S$ frame has velocity $-V$. Therefore, the transformation is 
\begin{equation}\label{eq4}
\left\{ \begin{array}{rcl}
          d\bar{x}^{0}\!\!\! & = & \,k(-V)\,\gamma (dx^{0} - \beta \, dx^{1}) \\
          d\bar{x}^{1}\!\!\! & = & \,k(-V)\,\gamma (-\beta \, dx^{0} + dx^{1}) \\
          d\bar{x}^{2}\!\!\! & = & \,k(-V)\, dx^{2}   \\
          d\bar{x}^{3}\!\!\! & = &\,k(-V)\, dx^{3}\,\,\, .
         \end{array} \right.
\end{equation}

\end{enumerate}

Thus, from Eqs.~(\ref{eq3}) and (\ref{eq4}), we have $k(V)k(-V)=1$.
Furthermore, on the frame $S$ set up a new coordinate system $\Xi$ which just reverses the positive directions of all three axes of the original coordinate system $X$. With respect to the new coordinate system $\Xi$, the state of motion of the particle is then 
$d\xi^{\alpha}$ of which $d\xi^{0}=dx^{0}$, and $d\xi^{i}=-dx^{i}$ $(i=1,2,3)$. Similarly, such a new coordinate system $\bar{\Xi}$ is set up on the frame $\bar{S}$.
With respect to the new coordinate systems $\Xi$ of the frame $S$, the frame $\bar{S}$ moves  with $-V$ relative to the frame $S$. Thus, the transformation is 
\begin{equation}\label{eq5}
\left\{ \begin{array}{rcl}
          d\xi^{0}\!\!\! & = & \,k(-V)\,\gamma (d\bar{\xi}^{0} - \beta \, d\bar{\xi}^{1}) \\
           d\xi^{1}\!\!\! & = &\,k(-V)\,\gamma (-\beta \, d\bar{\xi}^{0} + d\bar{\xi}^{1}) \\
          d\xi^{2}\!\!\! & = & \,k(-V)\, d\bar{\xi}^{2}   \\
          d\xi^{3}\!\!\! & = & \,k(-V)\,d\bar{\xi}^{3}\,\,\, .
         \end{array} \right.
\end{equation}
Since $d\xi^{0}=dx^{0}$, $d\xi^{i}=-dx^{i}$, $d\bar{\xi}^{0}=d\bar{x}^{0}$, and $d\bar{\xi}^{i}=-d\bar{x}^{i}$, from Eq.~(\ref{eq5}) we have
\begin{equation}\label{eq6}
\left\{ \begin{array}{rcl}
          dx^{0}\!\!\! & = & \,k(-V)\,\gamma (d\bar{x}^{0} + \beta \, d\bar{x}^{1}) \\
           dx^{1}\!\!\! & = & \,k(-V)\,\gamma (\beta \, d\bar{x}^{0} + d\bar{x}^{1}) \\
          dx^{2}\!\!\! & = & \,k(-V)\, d\bar{x}^{2}   \\
          dx^{3}\!\!\! &= & \,k(-V)\,d\bar{x}^{3}\,\,\, .
         \end{array} \right.
\end{equation}
Comparing between Eqs.~(\ref{eq3}), and (\ref{eq6}), we have $k(V)=k(-V)$.  As $k(V)>0$ and $k(V)k(-V)=1$ have been deduced, we obtain $k(V)=1$.
Finally, the relativistic transformation for $d x^{\alpha}$ in the velocity space is: 
\begin{equation}\label{eq7}
\left\{ \begin{array}{rcl}
          dx^{0}\!\!\! & = & \gamma (d\bar{x}^{0} + \beta \, d\bar{x}^{1}) \\
           dx^{1}\!\!\! & = & \gamma (\beta \, d\bar{x}^{0} + d\bar{x}^{1}) \\
          dx^{2}\!\!\! & = & \,d\bar{x}^{2}   \\
          dx^{3}\!\!\! & = & \, d\bar{x}^{3}\,\,\, .
         \end{array} \right.
\end{equation}

  The relativistic transformation of $d x^{\alpha}$ in the velocity space obtained herein Eq.~(\ref{eq7}) is just like the  Lorentz transformation in its differential form; we may name this relativistic transformation  the differential Lorentz transformation. However, it should be emphasized that here the infinitesimal quantities $d x^{\alpha}$ are not the differential of space-time coordinates $x^{\alpha}$.\cite{10}

It is interesting to note that the same deduction can be performed based on (1) the principle of relativity, and (2) a postulate of an invariant speed which is not necessarily equal to the speed of light.  With a postulate of an invariant speed $\sigma$, we obtain the same formulae as the transformation Eq.~(\ref{eq7}), but $dx^{0} \equiv \sigma \,dt$, and $\beta \equiv V / \sigma$ instead. Then, from this transformation, when $\sigma \rightarrow \infty$, we have the Galilean transformation in its differential form, 
\begin{equation}\label{eq8}
\left\{ \begin{array}{lcl}
         dt\!\!\! & = & \, d\bar{t} \\
           dx^{1}\!\!\! & = & \, V d\bar{t} + d\bar{x}^{1} \\
          dx^{2}\!\!\! & = & \,d\bar{x}^{2}  \\
          dx^{3}\!\!\! & = & \,d\bar{x}^{3}\,\,\, .
         \end{array} \right.
\end{equation}
When $\sigma = c$, we have the differential Lorentz transformation Eq.~(\ref{eq7}).

\section{Conclusions}

The meaning of the principle of relativity  is that all inertial frames are equivalent, that is, it is impossible by any means to distinguish whether or not inertial frames are intrinsically stationary. According to special relativity, all laws of physics are required to be form-invariant  under  Lorentz transformation of the space-time coordinates. Such a requirement on the laws of physics becomes practically equivalent to the principle of relativity. However,
this conventional interpretation of the principle of relativity is not the principle of relativity {\it per se}.\cite{11,12}

  Usually, the differential Lorentz transformation is considered to be equivalent to the usual Lorentz transformation of space-time coordinates. The  reasoning behind that is given in the following. The differential Lorentz transformation can be obtained by taking the differential of the usual Lorentz transformation.  Vice versa, the usual Lorentz transformation can be obtained by taking the integration of the differential Lorentz transformation with given initial conditions.  We here give reasons why these two transformations are not equivalent.  In order for these two transformations to be equivalent, it is required that the exact relationship of initial conditions of events between inertial frames must be attainable in any situation.   Nonetheless, there exists no law in special relativity that can in general provide the exact relationship of initial conditions of events between inertial frames. According to special relativity, the initial conditions of events between frames do not necessarily fulfill the requirements of form-covariance under the Lorentz transformation of space-time coordinates. Even more, according to the uncertainty principle of quantum mechanics, if one knows  exact information in the space of velocity, then one can not extract any information in the space of space-time coordinate, and vice versa. Based on the uncertainty principle, simultaneous transformation of both the space-time coordinates and velocities of events is untenable. 
The usual Lorentz transformation, which can simultaneously transform the exact values of the space-time coordinate and velocity, is incompatible with the uncertainty principle of quantum mechanics.\cite{13} The differential Lorentz transformation is a transformation in the velocity space; it is compatible with the uncertainty principle of quantum theory.


\begin{thebibliography}{}
\bibitem{8} "Let each system be provided with a rigid measuring-rod and a number of clocks, and let the two measuring-rods, and likewise all the clocks of the two systems, be in all respects alike." in reference 2.
\bibitem{10} As a familiar example, an infinitesimal heat  $dQ$ transferring to or from a system is not the differential of a non-existent heat function $Q$ of the system.
\end{thebibliography}
\end{document}